\documentclass{article}

\usepackage{arxiv}

\usepackage[utf8]{inputenc} 
\usepackage[T1]{fontenc}    
\usepackage{hyperref}       
\usepackage{url}            
\usepackage{booktabs}       
\usepackage{amsfonts}       
\usepackage{nicefrac}       
\usepackage{microtype}      
\usepackage{lipsum}		
\usepackage{graphicx}

\usepackage{doi}

\newtheorem{definition}{Definition}[section]

\usepackage{ifpdf}
\usepackage{amsmath}
\usepackage{algorithmic}

\usepackage{array}
\usepackage{fixltx2e}
\usepackage{url}

\usepackage[table]{xcolor}
\definecolor{lightgray}{rgb}{.9,.9,.9}
\definecolor{darkgray}{rgb}{.7,.7,.7}
\usepackage{ makecell}
\renewcommand\cellalign{lc}
\renewcommand\theadalign{bc}
\renewcommand\theadfont{\normalsize}
\usepackage{float}
\usepackage[most]{tcolorbox}
\usepackage{balance}
\usepackage{microtype}
\usepackage{xspace}
\usepackage[ruled,linesnumbered]{algorithm2e}
\usepackage{tabularx}
\usepackage{booktabs}
\usepackage{multirow}

\newcommand{\cs}{\textsf{CS score}\xspace}
\newcommand{\method}{\textsf{PPT}\xspace}
\newcommand{\methodnouq}{\textsf{PPT-NoUQ}\xspace}
\newcommand{\methodbuq}{\textsf{PPT-BUQ}\xspace}
\newcommand{\methodeuq}{\textsf{PPT-EUQ}\xspace}
\newcommand{\methodduq}{\textsf{PPT-DUQ}\xspace}
\newcommand{\methodnopl}{\textsf{PPT-NoPL}\xspace}
\newcommand{\sa}{\textsf{TTE analysis}\xspace}
\newcommand{\cbox}[2][!gray]{%
  \colorbox{#1}{\parbox{\dimexpr\linewidth-2\fboxsep}{#2}}%
  \vspace{-4mm}
}

\title{Pretrain, Prompt, and Transfer: Evolving Digital Twins for Time-to-Event Analysis in Cyber-physical Systems}

\author{ 
{Qinghua Xu} \\
	Department of Engineering Complex Software Systems\\
	Simula Research Laboratory\\
	Oslo, Norway \\
	\texttt{qinghua@simula.no} \\
	\And
	{Tao Yue} \\
	Department of Engineering Complex Software Systems\\
	Simula Research Laboratory\\
	Oslo, Norway \\
	\texttt{taoyue@gmail.com} \\
        \And
	{Shaukat Ali} \\
	Department of Engineering Complex Software Systems\\
	Simula Research Laboratory\\
	Oslo, Norway \\
	\texttt{shaukat@simula.no} \\
 \And
	{Maite Arratibel} \\
	Orona Group\\ 
	Hernani, Basque Country, Spain \\
	\texttt{marratibel@orona-group.com} \\
}

\renewcommand{\shorttitle}{\textit{arXiv} Template}

\hypersetup{
pdftitle={A template for the arxiv style},
pdfsubject={q-bio.NC, q-bio.QM},
pdfauthor={David S.~Hippocampus, Elias D.~Striatum},
pdfkeywords={First keyword, Second keyword, More},
}

\begin{document}
\maketitle

\begin{abstract}
	Cyber-Physical Systems (CPSs), e.g., elevator systems and autonomous driving systems, are progressively permeating our everyday lives. To ensure their safety, various analyses need to be conducted, such as anomaly detection and time-to-event analysis (the focus of this paper). Recently, it has been widely accepted that digital Twins (DTs) can serve as an efficient method to aid in the development, maintenance, and safe and secure operation of CPSs. However, CPSs frequently evolve, e.g., with new or updated functionalities, which demand their corresponding DTs be co-evolved, i.e., in synchronization with the CPSs. To that end, we propose a novel method, named \method, utilizing an uncertainty-aware transfer learning for DT evolution. Specifically, we first pretrain \method with a pretraining dataset to acquire generic knowledge about the CPSs, followed by adapting it to a specific CPS with the help of prompt tuning. Results highlight that \method is effective in time-to-event analysis in both elevator and ADSs case studies, on average, outperforming a baseline method by 7.31 and 12.58 in terms of Huber loss, respectively. The experiment results also affirm the effectiveness of transfer learning, prompt tuning and uncertainty quantification in terms of reducing Huber loss by at least 21.32, 3.14 and 4.08, respectively, in both case studies.
\end{abstract}

\section{Introduction}\label{sec:introduction}
Cyber-physical systems (CPSs) typically incorporate a cyber component, linking physical systems through a network. This combination of cyber and physical systems enables more intelligent and adept industrial applications, especially in crucial infrastructures such as transportation systems. However, the increasing complexity, heterogeneity, and constantly evolving nature of CPSs, brought about by introducing a rich array of functionalities, opens them up to significant threats and challenges. This often renders existing security and safety techniques ineffective, emphasizing the need to devise novel techniques to ensure the dependability of various CPS tasks.

Among these tasks, time-to-event (TTE) analysis~\cite{musa_deep_2023}, also known as survival analysis, is of great importance, as CPSs are characterized by the interaction of computational and physical processes, often facing uncertainty, and the reliability of the systems is of paramount importance. \sa allows for modeling and predicting the time until certain events occur, such as predicting the passenger waiting time in an elevator system and predicting time-to-collision in an autonomous driving system (ADS). \sa can also help to understand and quantify the reliability and operational resilience of the systems under different conditions or in response to different threats. Therefore, applying \sa in CPSs can facilitate CPS operators, and other relevant stakeholders, to take timely preventive measures, optimize resource allocation, etc., hence, making the systems safer and more efficient. 

\textcolor{black}{The literature on CPS behavior prediction also focuses on dynamic models, e.g., ordinary differential equations ~\cite{lee_cps_2010}. Such models, while offering systematic and mathematical ways to capture CPS dynamics, often face challenges in handling environmental uncertainties, adapting to real-time changes, and integrating different data sources, especially in the context of large and stochastic CPSs. Such CPSs, despite having well-defined models, exhibit complex, non-linear behaviors under certain conditions, hence it is challenging to accurately capture such behavior in a scalable manner with dynamic models alone. For instance, a new road sign can potentially induce uncertain behaviors of an ADS if the sign has not been defined in the dynamic model in advance. Various methods have been proposed to address these challenges of conventional dynamic models~\cite{li_recent_2020}. Among them,} Digital Twins (DTs) have gained substantial attention recently in the CPS domain. Early works ~\cite{eckhart_securing_2018} heavily rely on prior knowledge (e.g., rule-based models and domain expertise) to construct DTs, whereas data-driven DT construction is receiving increasing interest, due to the success of applying machine learning in software engineering. The efficacy of a DT functionality hinges on its synchronization with the real CPS, which inspires researchers and practitioners to create a DT that faithfully simulates the CPS. However, its continuous evolution, e.g., due to hardware or software updates, demands the evolution of its corresponding DT. One straightforward solution is to train a new DT from scratch with data collected from the updated CPS~\cite{xu_digital_2021,xu_digital_2023,xu_kddt_2023}. However, required data is not always guaranteed, e.g., in the case of a newly deployed elevator producing insufficient data for deep learning training.

To combat the data scarcity, in our prior work, we proposed RISE-DT ~\cite{xu_uncertainty-aware_2022}, an uncertainty-aware transfer learning method to evolve DT for industrial elevators. \textcolor{black}{Despite that transfer learning would add an extra uncertainty due to the disparity between knowledge domains, quantified as transferability, it can still provide noisy yet valuable information for performance improvement compared to training with only limited domain-specific data~\cite{nguyen_leep_2020}.} RISE-DT aims to transfer knowledge from the DT constructed for the source elevator system to a target (new) elevator system. Concretely, RISE-DT first employs uncertainty quantification (UQ) to select the most uncertain samples, which are the most informative samples as well since they tend to be close to the decision boundary. We then train a source DT and a target DT with these samples. The transfer learning process minimizes the conditional and marginal losses between the representations in the source and target DTs, allowing knowledge to be transferred across the domains. 

In this paper, we propose \method to extend RISE-DT. Our key contributions are three-fold. First, we improve the RISE-DT's performance by employing prompt tuning, which has emerged as an effective method for tuning pretrained models, especially for large language models, to downstream tasks~\cite{brown_language_2020}. 
Second, comparing with RISE-DT, we study two more UQ methods, i.e., Bayesian and ensemble methods, to select the most suitable one for \sa. Third, we introduce an ADS dataset to demonstrate the generalizability of \method. Hence, we evaluate \method's application in two domains: elevator systems (vertical transportation) and ADS (horizontal transportation). Results show that \method is effective in \sa in both case studies, averagely outperforming the baselines by 7.31 and 12.58, respectively, regarding Huber loss. We also dissect the individual contribution of each subcomponent in \method and find that prompt tuning, UQ, and transfer learning are all effective and efficient.

\section{Background and Definitions}
\label{sec:background}
\subsection{CPS evolution}
A typical CPS architecture has physical and cyber elements, the symbiotic relationship facilitated through a feedback loop, incorporating sensors, actuators, communication networks, and computational units. The advances in developing these components, particularly computational units, have spurred the widespread adoption of CPSs in our daily lives. 

\textbf{CPS evolution} is often triggered by internal changes, such as upgrading old or introducing new CPS functionalities~\cite{t_yue_evolve_2023}. CPS behaviors are also closely intertwined with their operating environment. Such an operating environment can be very dynamic and uncertain, e.g., the driving environment of autonomous vehicles, which subsequently influences their decision-making at runtime. Therefore, we posit that a CPS should be studied along with its operating environment. Formally, we define a subject system $\Sigma$ comprising the CPS $\Psi$ and its environment $\Phi$ below:
\begin{equation}
    \label{eq:cps}
    \Sigma: \Psi \rightleftharpoons \Phi 
\end{equation}
Correspondingly, the evolution of the CPS is, thus, defined in Equation \ref{eq:evolution}, where $\Sigma_S$ and $\Sigma_T$ are the source and the target systems of the evolution, and $\Delta \Psi_S$ and $\Delta \Phi_S$ represent changes in the CPS and the environment, respectively. 
\begin{equation}
    \label{eq:evolution}
    \Sigma_S \xrightarrow[\Delta \Psi_S]{\Delta \Phi_S} \Sigma_T
\end{equation}

\subsection{\sa in industrial elevators and ADSs}

\textbf{TTE analysis} tasks, in general, can be described as a 4-tuple: $\langle \Sigma, \mathcal{D}, \mathcal{E},\tau \rangle$, where $\Sigma, \mathcal{D}, \mathcal{E}, \tau$ represent the subject system, dataset, events of interest, and time interval. \sa analyses dataset $\mathcal{E}$ collected from $\Sigma$ to predict time interval $\tau$ after which event $\mathcal{E}$ will occur (Equation \ref{eq:sa}). 
\begin{equation}
    \label{eq:sa}
    f: \mathcal{D}\mapsto {\tau_\mathcal{E}}
\end{equation}

\textbf{Industrial elevators} are vertical transportation systems for buildings and are essential for modern urban lives. The cyber aspect of an elevator, such as the control algorithm, is encapsulated in the elevator software, while the physical components, including motors, brakes, and cables, facilitate the movement of the elevator. Typically, each building has multiple elevators deployed and controlled by dedicated controllers, which are connected to a \textit{traffic master} with a \textit{dispatcher} (i.e., software) scheduling elevator operations to optimize the Quality of Services (QoS) to deliver. A common \sa is about predicting the waiting time of each passenger based on information such as arrival floor, destination floor, weight, and capacity. We formally define the elevator \sa task as in Definition \ref{def:elevator}.
\begin{definition}[Industrial Elevator TTE Analysis]  
    \label{def:elevator}
    \begin{equation}
    \begin{aligned}
        \Sigma^E&\mapsto{\text{An elevator system and its environment}}\\
         \mathcal{D}^E&\mapsto{\text{A sequence of passenger information}}\\
         \mathcal{E}^E&\mapsto{\text{Passengers arrive at their destinations}}\\
         \tau^E&\mapsto{\text{Estimated arrival time}}  
    \end{aligned}
\end{equation}
\end{definition}

Elevator dispatchers across buildings and their usage patterns depend on traffic factors such as building type, time of day, and day of the week (known as traffic template). Thus, both elevator dispatchers and their environment evolve, which affects \sa's performance. We define the evolution directions in Definition \ref{def:ele_evo}, where $\Delta \Psi^E_S$ represents CPS changes, that is, dispatcher version changes, and $\Delta \Phi^E_S$ denotes environment changes, i.e., traffic template changes. 
 \begin{definition}[Industrial Elevator Evolution]
     \label{def:ele_evo}
     \begin{equation}
    \Sigma^E_S \xrightarrow[\Delta \Psi^E_S]{\Delta \Phi^E_S} \Sigma^E_T
\end{equation}
 \end{definition}

\textbf{ADSs}, as another type of CPSs, have various sensors, e.g., optical and thermographic cameras, radar, and lidar. Their cyber part is mainly responsible for planning and controlling the vehicles' behavior. Specifically, an ADS uses sensors to perceive its environment (e.g., road conditions and other vehicles), which are then utilized to guide the ADS's decision-making, e.g., determining an appropriate navigation path and formulating strategies to manage traffic controls (e.g., stop signs) and obstacles. \sa such as predicting the time to a potential collision (known as time-to-collision) can facilitate the ADS to make well-informed decisions, which is defined as in Definition \ref{def:ads_tte}.
\begin{definition}[ADS TTE Analysis]
    \label{def:ads_tte}
    \begin{equation}
    \label{eq:elevator}
    \begin{aligned}
        \Sigma^A&\mapsto{\text{An ADS and its running environment}}\\
         \mathcal{D}^A&\mapsto{\text{A sequence of vehicle and environment properties}}\\
         \mathcal{E}^A&\mapsto{\text{Vehicle collisions}} \\
         \tau^A&\mapsto{\text{Estimated collision time}}  
    \end{aligned}
\end{equation}
\end{definition}
\textcolor{black}{ADS behaviors differ under varying internal functionalities and driving conditions, including weather conditions and behaviors of nearby vehicles. In this work, we only concern the evolution of ADSs under different driving conditions as depicted in Definition \ref{def:ads_evo}, an area widely studied in ADS testing~\cite{zhang_finding_2023}. The DeepScenario dataset we used in this study, unfortunately, does not consider ADS's internal functionalities and, hence, cannot be used to study the functionality evolution as in the elevator case study.}
%
%
\begin{definition}[ADS Evolution]
    \label{def:ads_evo}
     \begin{equation}
    \Sigma_S^A \xrightarrow{\Delta \Phi^A_S} \Sigma_T^A
\end{equation}
\end{definition}

\subsection{Digital Twin}
\label{subsec:dt}
\begin{figure}[!htb]
\centering
\includegraphics[width=0.5\columnwidth]{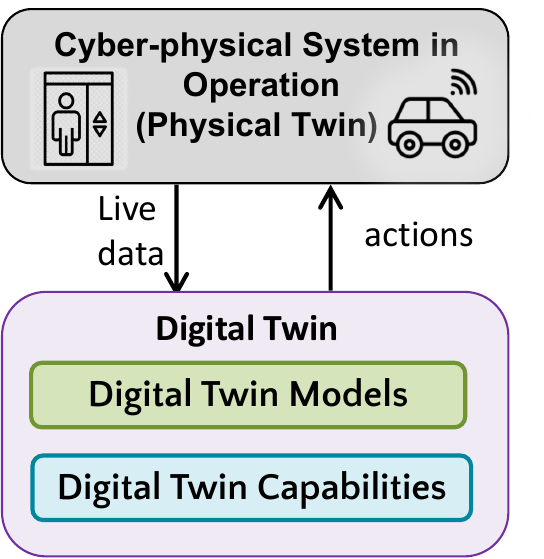}
\caption{Digital Twin for Cyber-physical System}
\label{fig:dt}
\end{figure}

El Saddik ~\cite{a_el_saddik_digital_2018} defined DT as a digital replica of a physical entity. Yue et al.~\cite{yue_understanding_nodate} extended the definition and proposed a conceptual model (Figure \ref{fig:dt}), which defines a CPS (e.g., an elevator system or an ADS) as a physical twin. Its DT comprises a Digital Twin Model (DTM) and a Digital Twin Capability (DTC). The DTM is a digital representation of the CPS, including heterogeneous models corresponding to various components, e.g., software, hardware, and communication. The DTC is the DT's functionality, e.g., detecting anomalies and preventing failures. \method adopts this conceptual model for DT construction and evolves the DTM and DTC. 

\subsection{Uncertainty Quantification}
UQ is intensively studied in both academia and industry, underpinning numerous applications such as trustworthy decision-making~\cite {abdar_review_2021} and software risk analysis~\cite{lin_uncertainty_2021}. In our context, we use UQ to select the most uncertain samples $U$ from a dataset $\mathcal{D}$. To that end, UQ assigns an uncertainty score $\xi$ to each sample $x\in \mathcal{D}$. We have defined a comprehensive UQ metric \cs in our prior work ~\cite{xu_uncertainty-aware_2022}. In this paper, we further investigate UQ by adding two mainstream UQ approaches: Bayesian and ensemble UQ. Let $\mathcal{M}_{I}$ be an indicator model which assesses the uncertainty of each sample and shares the same structure as the DTM (Section \ref{subsubsec:dt_component}). We introduce each UQ approach below.

\cs combines the calibration and sharpness metrics. Calibration represents the consistency between the prediction distribution and the observation, while sharpness assesses the concentration of the prediction distribution~\cite{gneiting_probabilistic_2007}. By combining them with a weighted sum parameterized by $\lambda$ (decided empirically), we follow~\cite{xu_uncertainty-aware_2022} and define the comprehensive uncertainty metric \textit{CS score} $\xi^{cs}$ (Equation \ref{eq:ut_metric}). \textcolor{black}{Parameter $\lambda$ signifies the relative importance between calibration and sharpness. If $\lambda$ is high, the metric places more emphasis on the calibration and vice versa.}
    \begin{equation}
        \label{eq:ut_metric}
        \xi_i^{ut}=\lambda c(x_i)+ (1-\lambda) s(x_i)
    \end{equation}

\textit{Bayesian Method} probabilistically interprets predictions, which can be leveraged to derive UQ metrics. One popular Bayesian UQ method in neural network models is the Monte Carlo (MC) dropout, which randomly sets the activation of neurons to 0 with a fixed probability for a subset of layers, resulting in a set of indicator models with dropout $\{\mathcal{M}^B_d\}_{d=1}^{N_B}$, where $N_B$ is the number of indicator models. 
Each indicator model makes an individual prediction, and we define the uncertainty of each sample as the standard deviation of these predictions as in Equation \ref{eq:bm}. 
    \begin{align}
        \label{eq:bm}
        \Bar{y}_i &=\frac{1}{N_B}\sum_{d=1}^{N_B}\mathcal{M}_d^B(x_i)\\
        \xi_i^{bm} &=\sqrt{\frac{1}{N_B}\sum_{d=1}^{N_B}(\mathcal{M}_d^B(x_i-y_i)^2} 
    \end{align}

\textit{Ensemble Method} is a commonly used technique in machine learning to combat overfitting by training multiple models with different configurations simultaneously. Building on the idea of ensemble learning, ensemble UQ divides the dataset into $N^E$ subsets and trains a distinct indicator model $M^E$ on each subset. Similar to the MC dropout approach, each indicator model generates predictions independently, and the uncertainty of each sample $x_i$ is determined by calculating the standard deviation of the predictions, as shown in Equation \ref{eq:em}.
    \begin{align}
        \label{eq:em}
        \Bar{y}_i &=\frac{1}{N_E}\sum_{d=1}^{N_E}\mathcal{M}_d^E( x_i)\\
        \xi_i^{em} &=\sqrt{\frac{1}{N^E}\sum_{d=1}^{N_E}(\mathcal{M}_d^E(x_i-y_i)^2} 
    \end{align}

\section{Approach}\label{sec:method}

\begin{figure*}

    \centering
    \includegraphics[width=1.7\columnwidth]{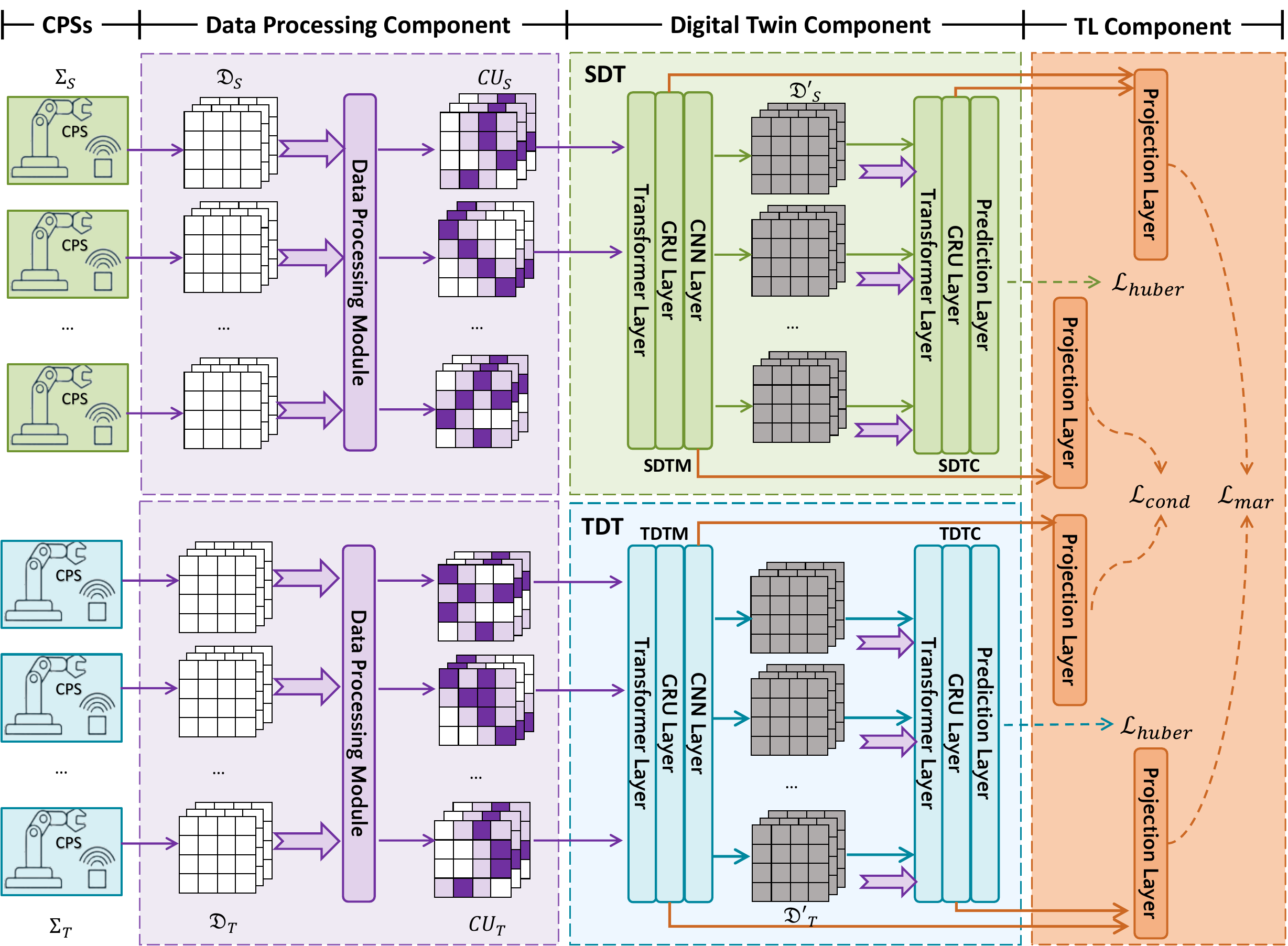}
    \caption{Overall architecture of  \method }
    \label{fig:architecture}
\end{figure*}

\method is a closed-loop deep learning approach, which requires training on the relevant dataset. We introduce the architecture of \method in Section \ref{subsec:structure} and the training method in Section \ref{subsec:training}.

\subsection{Overall Architecture}\label{subsec:structure}
Figure \ref{fig:architecture} depicts the overall architecture of \method, comprising the \textit{Data Processing Component}, the \textit{Digital Twin Component} and the \textit{Transfer Learning Component} (denoted as TL component in the figure). Let the source and target subject systems be $\Sigma_S$ and $\Sigma_T$ and their corresponding datasets be $\mathcal{D}_S$ and $\mathcal{D}_T$. The \textit{data processing component} takes as input $\mathcal{D}_S$ and $\mathcal{D}_T$ and selects the most uncertain contextualized samples $CU_S$ and $CU_T$. The source and target \textit{digital twin components} use these samples to construct a source DT (SDT) and a target DT (TDT). Finally, the \textit{Transfer Learning Component} transforms hidden representations in SDT and TDT into shared intermediate spaces, which signify the shared knowledge between the source and target domain. 


\subsubsection{Data Processing Component}\label{subsubsec:data_process_component}
The data processing component takes source and target data ($\mathcal{D}_S$ and $\mathcal{D}_T$) as input and selects the most uncertain samples with their context information using the data processing module. According to \textit{Information Theory}, higher uncertainty entails richer information~\cite{klir_uncertainty_2006}. Hence, machine learning models can greatly benefit from training with more informative samples~\cite{y_yang_active_2016}.

\begin{figure}
    \centering
    
    \includegraphics[width=0.7\columnwidth]{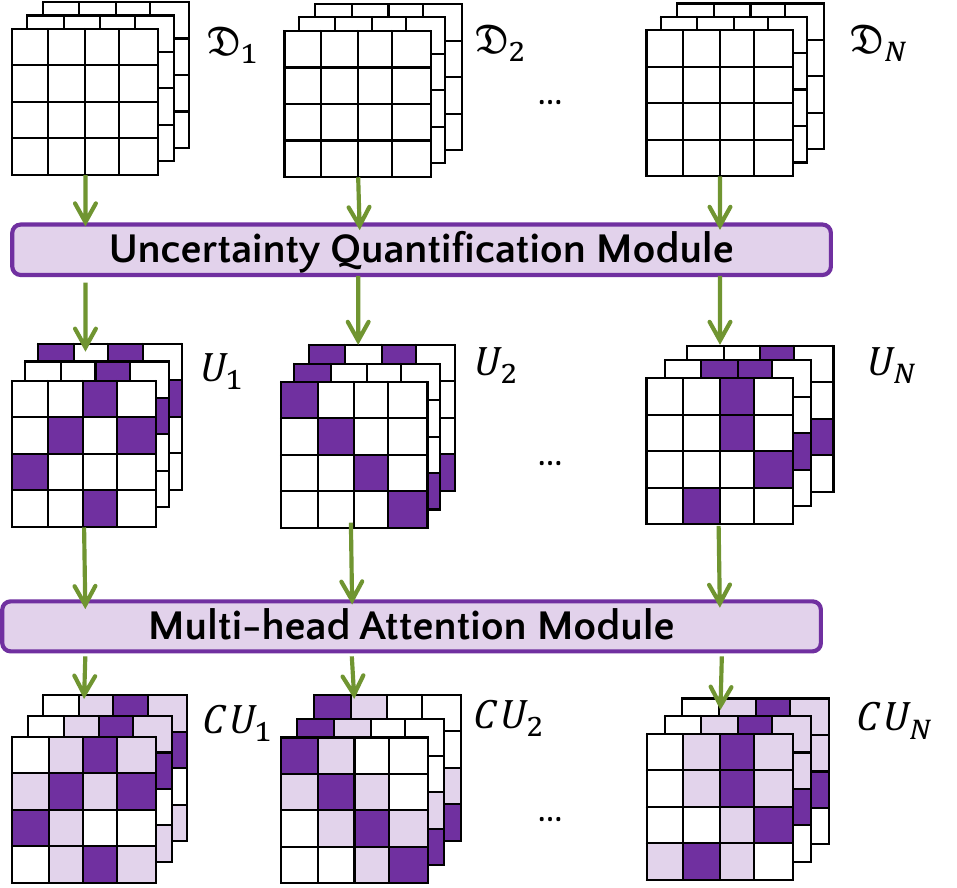}
    \caption{\textcolor{black}{Structure of the data processing module. Dark purple squares represent the selected most uncertain samples, light purple squares signify their contexts, and white squares denote unselected samples.}}
    \label{fig:data_process}
\end{figure}

We illustrate the details of the data processing module in Figure \ref{fig:data_process}. 
Given datasets $\mathcal{D}_1,\mathcal{D}_2,...,\mathcal{D}_N$, we utilizes an \textit{uncertainty quantification module} to select the most uncertain samples $U_1, U_2, ..., U_N$. However, their context information is also critical for \sa, since each sample is dependent on previous samples and influences future ones. Picking out the most uncertain samples has the possibility of losing the context information. Hence, we use a \textit{multi-head attention module} to calculate representations $CU_1, CU_2, ..., CU_N$ that fuse information from both the sample itself and its context. 
 
\textbf{Uncertainty Quantification Module} ranks and selects the most uncertain samples. In this study, we explore three UQ methods: \cs, Bayesian and ensemble UQ methods, and select the most suitable one. UQ assigns an uncertainty score $\xi_i \in \Xi$ to each sample $x_i\in \mathcal{D}$. We then rank all samples based on their scores and select the top $K$ ranked samples for transfer learning (Equation \ref{eq:rank}).
\begin{equation}
    \label{eq:rank}
    U=topK(\mathcal{D},key=\Xi)
\end{equation}

\textbf{Multi-head Attention Module}.
\textcolor{black}{
CPS data is often contextual, meaning that considering only uncertain samples in isolation could potentially harm the model's performance. To preserve the contextual information, we employed multi-head self-attention (MHSA)~\cite{vaswani_attention_2017}. MHSA first projects the input data into a hidden space, where each vector contains information about both the input itself and its surrounding context. Then, MHSA enables the model to incorporate contextual information based on their importance and yields contextualized representations $CU$ as in Equation \ref{eq:mhsa}.} 
\textcolor{black}{\begin{equation}
    \label{eq:mhsa}
    CU=MHSA(U)
\end{equation}}


\subsubsection{Digital Twin Component}\label{subsubsec:dt_component}
In \method, a generic DT has a DTM simulating the physical twin and a DTC with functionalities, e.g., waiting time prediction.
\textbf{DTM} aims to approximate the underlying distribution of input data ($\mathcal{D}$) with three layers: the transformer, GRU and prediction layers, which extract features from each sample, capture temporal features and project the intermediate representation vectors into the sample space to predict the next sample with CNN, respectively.

\textit{The transformer layer} takes the contextualized uncertain samples $CU$ as input and feeds them into a stack of $L$ MHSA modules. \textcolor{black}{Transformers are known for their ability to capture long-range dependencies in data sequences, making them well-suited for modeling complex relationships within time-series data.} Each MHSA module takes the output of the prior MHSA module as input (Equation \ref{eq:transformer}).
\begin{equation}
    \label{eq:transformer}
    \begin{aligned}
    CU^M_1&=CU\\
        CU_L^M&=MHSA(CU^M_{L-1})
    \end{aligned}
\end{equation}

\textit{The GRU layer} takes the output of the transformer layer as input. \textcolor{black}{GRU, as a typical sequence model, is adept at handling sequence data while demonstrating its advantages in training efficiency, handling vanishing gradient problems and avoiding overfitting, compared to other sequence models such as Recurrent Neural Networks and Long short-term Memory Networks. } Concretely, for each data sample $x_i\in CU^M_L$, the GRU layer computes the hidden representation ($H^{G}_{M}$) using Equation \ref{eq:gru}.
    \begin{align}
        \label{eq:gru}
        z_t &=\sigma_g(W_z x_t + U_z h_{t-1}+b_z)\\
        r_t &=\sigma_g(W_r x_t + U_r h_{t-1}+b_r)\\
        \hat{h}_t &=\phi_h(W_h x_t+ U_h(r_t \cdot h_{t-1})+ b_h)\\
        H^{G}_{M}[t] &=(1-z_t) \cdot h_{t-1}+ z_t \cdot \hat{h}_t
    \end{align}
\textcolor{black}{Specifically, the GRU layer uses update and reset gates, denoted by $z_t$ and $r_t$, respectively, to modulate the flow of information in the sequence. The update gates decide how much historical information to keep while the reset gates control how much new information to integrate. 
} 

\textit{The CNN layer} predicts the next state vector for the subject system, where $S$ is the size of the state vector. \textcolor{black}{We use a CNN layer to capture fine-grained patterns and spatial features of each data sample. This layer allows the model to focus on specific segments of the data where localized features may be crucial for making accurate predictions.} Each of its dimensions is a continuous scalar value. However, direct training with continuous labels can result in overfitting issues with insufficient data. To overcome this, we discretize continuous scalar values into 10 categories, thereby transforming these continuous prediction tasks into classification tasks. The core operation of the CNN layer is the kernel convolution, which calculates the probability $P_{i,j}$ for $i$th label on the $j$th dimension (Equation \ref{eq:dtm_prob}). $\mathcal{K}$ denotes the convolution kernel. 
 \begin{equation}
     \label{eq:dtm_prob}
     P_{i, j}=\sum_{m}\sum_{n} X_{i-m, j-n} \cdot \mathcal{K}_{m, n}
 \end{equation}
We then predict the next data samples by assigning labels to each dimension with the highest probabilities (Equation \ref{eq:dtm_pred}).
\begin{equation}
    \label{eq:dtm_pred}
    CU^M=argmax(P_{i,j})
\end{equation}

    
\textbf{DTC} performs \sa. Figure \ref{fig:architecture} presents the design of the source DTC (SDTC) and target DTC (TDTC), which have identical architectures. Below, we only illustrate SDTC's architecture for brevity. 
SDTC combines real data $\mathcal{D}_S$ and predicted data $\mathcal{D}_S'$ as input and feeds it into three layers sequentially: the transformer, GRU, and prediction layers.

\textit{The transformer layer} concatenates the real sample $CU$ and predicted sample $CU^M$ and feeds it to a size-L stack of MHSA modules computed recursively (Equation \ref{eq:dtc_transformer}).  
\begin{equation}
    \label{eq:dtc_transformer}
    \begin{aligned}
    CU^C_1&=concat([CU,CU^M])\\
    CU^C_L&=MHSA(CU^C_{L-1}))
    \end{aligned}
\end{equation}

\textit{The GRU layer} captures the dependency between the current and previous inputs (Equation \ref{eq:dtc_gru}). Its detailed structure is described in Equation \ref{eq:gru}, where $H^{G}_{C}$ denotes the output of DTC's GRU layer. 
    \begin{equation}
    \label{eq:dtc_gru}
        H^{G}_{C}=GRU(CU^C_L)
    \end{equation}
%
 %
 
    \textit{The prediction layer} transforms the intermediate representations into the estimated time (e.g., waiting time in the elevator case study and time-to-collision in the ADS case study) as in Equation \ref{eq:dtc_pred}, where $W_{\tau} $ and $b $ are weight matrices.
    \begin{equation}
        \label{eq:dtc_pred}
       \hat{\tau}=W_{\tau}^T H^{G}_{C}+b_{\tau}
    \end{equation}

\subsubsection{Transfer Learning Component}\label{subsubsec:tl_component}
As shown in Figure \ref{fig:architecture}, \method uses a projection layer to map the hidden representations in SDT and TDT to shared spaces. 
\textit{The projection layer} first uses a linear transformation to map the hidden representations ($H$) to representations $H^P$ in the shared spaces and benefits from an activation function $tanh$ to add non-linearity (Equation \ref{eq:projection_layer}).
\begin{equation}
    \label{eq:projection_layer}
        H^P=tanh(W_P H+ b_P)
\end{equation}
Then, transfer learning aligns SDT and TDT in the intermediate spaces to reduce marginal and conditional losses.

\textcolor{black}{\textit{Marginal loss} is computed using the Kullback-Leibler (KL) divergence, a statistical measure for the difference between two probability distributions. We apply this to the hidden layer representations of SDT and TDT.} We first project the GRU layer outputs of SDTM, SDTC, TDTM and TDTC ($H^{G}_{SM}$, $H^{G}_{SC}$, $H^{G}_{TM}$, and $H^{G}_{TC}$) into an intermediate space with Equation \ref{eq:projection_layer}, yielding $H^{PG}_{SM}$, $H^{PG}_{TM}$, $H^{PG}_{SC}$ and $H^{PG}_{TC}$, respectively. We then calculate the marginal loss in the intermediate space as in Equation \ref{eq:ml_loss}. 
\begin{equation}
    \label{eq:ml_loss}
    \begin{aligned}
        \mathcal{L}_{mar}^{M} & =\sum_t{H^{G}_{SM}[t]\cdot logH^{G}_{TM}[t]}\\
    \mathcal{L}_{mar}^{C} & =\sum_t{H^{G}_{SC}[t]\cdot logH^{G}_{TC}[t]} 
    \end{aligned}
\end{equation}
\textcolor{black}{Marginal loss measures the divergence in the probability distributions of these components to ensure that the student and teach models have an aligned understanding of data.}

\textit{Conditional loss} $\mathcal{L}_{cond}^{M}$ is calculated between the prediction layer representations of SDTC and TDTC. We first transform the output of SDTC and TDTC into an intermediate space as in Equation \ref{eq:cond_project}.
\begin{equation}
    \label{eq:cond_project}
    \begin{aligned}
    P_S^P&=proj(P_S)\\
    P_T^P&=proj(P_T)
\end{aligned}
\end{equation}
We then calculate the Maximum Mean Discrepancy (MMD) in the intermediate space (Equation \ref{eq:cond_loss}). \textcolor{black}{MMD, as a non-parametric method, compares the similarity of two probability distributions, which is effective for transfer learning~\cite{farahani_brief_2021}.}
\begin{equation}
    \label{eq:cond_loss}
    \mathcal{L}_{cond}^{M}=||\frac{1}{n^s}\sum_{i=1}^{{n^s}}{P_S^P[i]}+\sum_{i=1}^{{n^t}}{P_T^P[i]} \frac{1}{n^t}||
\end{equation}
\textcolor{black}{The conditional loss assesses how closely the student model's prediction is to the teacher model's. It ensures the student model not only replicates the teacher model's prediction but is sensitive to the prediction distribution.}

\label{subsec:training}
\subsection{Training Process of \method}  
This process includes the pretraining phase (Section \ref{subsubsec:pp}) and prompt tuning (Section \ref{subsubsec:pl}) phase. The former induces better initialization for the model parameters, while the later quickly adapts the model to the target subject system. 

\subsubsection{Pretraining Phase}\label{subsubsec:pp}
Neural network methods, including \method, tend to be trapped in local optimal easily. Pretraining on large datasets can alleviate this issue by inducing the optimizer more towards the global optimal. In this phase, we aim to find the optimal parameters for \method, as described in Algorithm \ref{alg:pt}. We collect source and target dataset pairs $\langle\mathcal{D}_S^{pre}, \mathcal{D}_T^{pre}\rangle$ and output the pretrained $SDT^{pre}$ and $TDT^{pre}$ (Lines 3-4). In each pair, we first perform UQ to select the $K$ most uncertain samples for transfer learning (Lines 5-10). SDT and TDT take these samples as input and make predictions (Lines 11-14). We calculate the marginal loss and conditional loss (Lines 15-16) to accomplish transfer learning. Additionally, we calculate the Huber loss between the predicted TTE ($\tau'_S$ and $\tau'_T$) and the real TTE ($\tau_S$ and $\tau_T$). Minimizing Huber loss (Section \ref{subsec:metric}) can induce the DTC to make more accurate \sa. The last step of Algorithm \ref{alg:pt} minimizes all losses by adjusting the model parameters (Line 19).

\begin{algorithm}
\label{alg:pt}
\caption{\textcolor{black}{Pretraining Phase of \method}}
\footnotesize	
\SetAlgoLined
\KwIn{$\Sigma^{SPre}$ and $\Sigma^{TPre}$: source and target subject systems; $N$: number of source and target system pairs.}
\KwOut{$SDT^{pre}$ and $TDT^{pre}$: the pretrained source and target DTs.
}
$SDT^{Pre},TDT^{Pre}=initialize()$\;
 \For {i in 1:N}
 {
$\mathcal{D}^S_i$=collect\_from($\Sigma^{SPre}_i$)\;
$\mathcal{D}^T_i$=collect\_from($\Sigma^{TPre}_i$)\;
    \tcc{Data processing}
$\omega^S$=UQ($\mathcal{D}^{SPre}_i$)\;
$\omega^T$=UQ($\mathcal{D}^{TPre}_i$)\;
$U^S=topK(\mathcal{D}^{SPre}_i,key=\omega_i^S$)\;
$U_T=topK(\mathcal{D}^{TPre}_i,key=\omega_i^T$)\;
$CU^S$=MHSA($U^S$)\;
$CU^T$=MHSA($U^T$)\;
\tcc{Train $SDT$ and $TDT$ with transfer learning}
$H_{SM}^{G},P_S,CU_{SM}=SDTM(CU_S)$\;
$H_{TM}^{G},P_T,CU_{TM}=SDTM(CU_T)$\;
$H_{SC}^{G},\hat{\tau}_S=SDTC(CU_S,CU_{SM})$\;
$H_{TC}^{G},\hat{\tau}_T=TDTC(CU_T,CU_{TM})$\;

$\mathcal{L}_{mar}=mar(H_{SM}^{G},H_{TM}^{G})+mar(H_{SC}^{G},H_{TC}^{G})$\;
$\mathcal{L}_{huber}=huber(\hat{\tau}_S,\tau_S)+huber(\hat{\tau}_T,\tau_T)$\;
$\mathcal{L}_{cond}=conditional(P_S,P_T)$\;
$\mathcal{L}=\mathcal{L}_{huber}+\mathcal{L}_{cond}+\mathcal{L}_{mar}$\;
$minimize(\mathcal{L}, SDT^{Pre},TDT^{Pre})$\;
 }
\end{algorithm}

\subsubsection{Prompt Tuning Phase}\label{subsubsec:pl}

\begin{figure}
    \centering
    \includegraphics[width=0.48\textwidth]{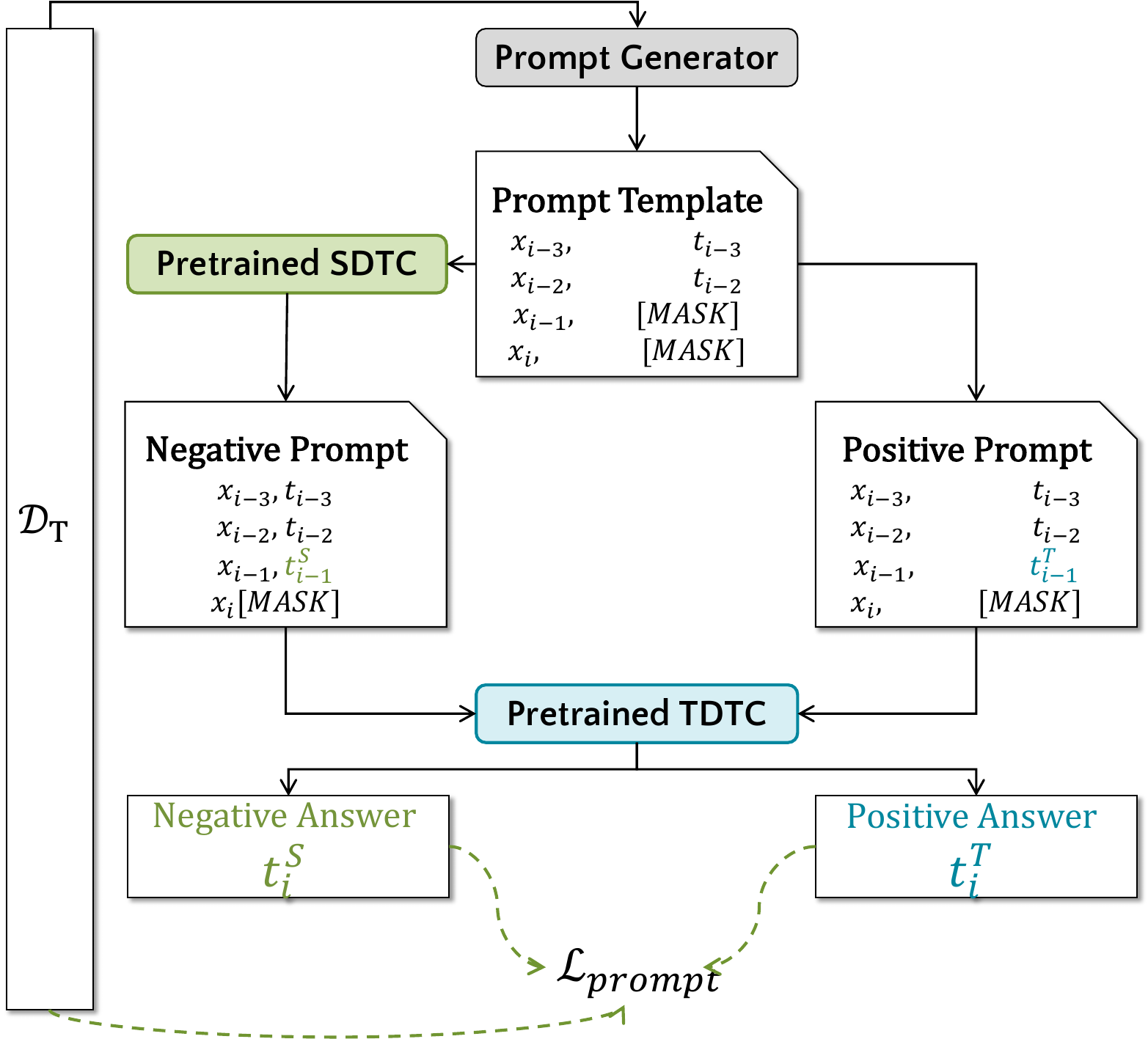}
    \caption{\method's prompt tuning. Note that we only display components related to prompt tuning and omit the detailed DT structure and UQ strategies for brevity since it is identical to those in the pretraining phase.}
    \label{fig:pl}
\end{figure}
The pretrained DTs are trained on the pretraining datasets $\mathcal{D}_{pre}$, which does not include the dataset collected from the target subject system $\mathcal{D}_T$. 
To acquire expertise in the target subject system $\Sigma$, supervised learning with a sufficient dataset collected from $\Sigma$ is required. For this purpose, we employed the fine-tuning technique in our previous work RISE-DT \cite{xu_uncertainty-aware_2022}. However, recent research has shown that prompt tuning can be even more effective \cite{liu_p-tuning_2022}. Hence, in \method, we employ prompt tuning to enhance its overall performance. Prompt tuning involves designing prompts to test a pretrained model's ability to distinguish between the source and target domain data. The feedback from the test helps the model learn more about the salient features in the target domain, potentially leading to improved performance.
A typical prompt tuning phase has three steps: prompt template designing, answer generation, and answer mapping \cite{liu_pre-train_2021}, as depicted in Figure \ref{fig:pl}. 

\textbf{Step 1: Prompt template designing.} 
Cloze-style prompts are a well-studied technique, where certain parts of the data are masked, leaving a blank for the model to fill in. At each time point $i$, we collect a L-lengthened sequence of CPS data $x_i,x_{i-1}, x_{i-2}, ..., x_{i-L+1}$. We generate a template by masking the time interval $\tau_i$ for the current time point $i$ and the previous one $i-1$ (denoted as "\textit{[ MASK ]}" in Figure \ref{fig:pl}). Using this template, we generate a positive prompt, where we fill in the true time interval $\tau^T_{i-1}$ for time point $i-1$ from the target domain (denoted as \textit{$\tau_T$} in Figure \ref{fig:pl}), and a negative prompt, where we fill in the same blank with the time interval predicted with SDTC (denoted as \textit{$\tau^S_{i-1}$} in Figure \ref{fig:pl}). 

\textbf{Step 2: Answer generation.} We ask the target DTC (TDTC) to fill in the blank in both positive and negative prompts. We hypothesize that SDTC captures knowledge about the source domain, while TDTC captures knowledge about the target domain. Therefore, we expect to make accurate predictions on the positive prompt while making noticeable errors when predicting the negative prompt, as it has been "tampered" by SDTC. As shown in Figure \ref{fig:pl}, TDTC fills in the two prompts, yielding a positive answer (denoted as \textit{Positive Answer} in Figure \ref{fig:pl}) and a negative answer (denoted as \textit{Negative Answer} in Figure \ref{fig:pl}). We calculate the Huber loss by comparing these two answers with the actual time interval $W\tau_i$ using Equation \ref{eq:loss_prompt}. Note that we reverse the sign of the Huber loss for the negative prompt by multiplying it by $-1$ because we assume a well-adapted TDTC should be able to distinguish between source and target domain data. This approach helps the model learn the salient features of the target domain and improves its performance, for instance, in predicting waiting times for elevator passengers in the target domain for our elevator case study and predicting the time-to-collision in the target domain for the ADS case study.
\begin{equation}
    \label{eq:loss_prompt}
    \mathcal{L}_{prompt}=(\tau-\tau_+)^2-(\tau-\tau_-)^2
\end{equation}

\textbf{Step 3: Answer mapping.} Compared to fine-tuning, prompt tuning can reduce or even obviate the need for extra model extensions. In prompt tuning, downstream task predictions are acquired by mapping prompt answers to the prediction space. In our context, we do not need to perform such mapping since the positive prompt prediction can be considered as \sa directly.

Algorithm \ref{alg:pl} describes the prompt tuning phase in pseudo-code. For each time point $i$, we consider not only the current data point but also history data points within a time window of $K$. We generate positive and negative prompts with the help of the prompt generator (Lines 4-6). SDT and TDT make predictions with $\mathcal{D}_S$ and the latest $\mathcal{D}_T$ data, respectively (Lines 7-10). In Lines 11-13, TDTC fills in the positive prompt and negative prompt. We calculate the prompt loss function as in Equation \ref{eq:loss_prompt}, and optimize the parameters of DTs stochastically (Lines 14-15).

\begin{algorithm}
\label{alg:pl}
\SetAlgoLined
\caption{Prompt Tuning Phase of \method}
\footnotesize	
\KwIn{$\mathcal{D}_S$ and $\mathcal{D}_T$: source and target subject system data; $SDT_{pre}$ and $TDT_{pre}$: pretrained source and target DTs; $\omega$: history window size }

\KwOut{$TDT$: Trained target DT}
$TDT=initialize()$\;
$i=0$\;
\While{$i<length(\mathcal{D}_T)$}
{
\tcc{Generate prompt and prompt answer}
$\rho=PromptGenerator(\mathcal{D}_T[i-\omega:i+1])$\;
$\rho_+=fill\_in(\rho,\mathcal{D}_T[i])$\;
$\rho_-=fill\_in(\rho,SDTC(\mathcal{D}_T[i-\omega:i]))$\;
\tcc{Train SDT and TDT with transfer learning and prompt tuning}
$H_{SM}^{G},P_S,CU_{SM}=SDTM(CU_S)$\;
$H_{TM}^{G},P_T,CU_{TM}=SDTM(CU_T)$\;
$H_{SC}^{G},\hat{\tau}_S=SDTC(CU_S,CU_{SM})$\;
$H_{TC}^{G},\hat{\tau}_T=TDTC(CU_T,CU_{TM})$\;
$\tau_+=TDTC(\rho_+)$\;
$\tau_-=TDTC(\rho_-)$\;
$\tau=\mathcal{D}_T[i]$\;
$\mathcal{L}_{prompt}=(\tau-\tau_+)^2-(\tau-\tau_-)^2$\;
$minimize(\mathcal{L}_{prompt}, TDT)$\;
$i=i+1$\;
}
\end{algorithm}


\section{Experiment Design}
\label{sec:experiment_design}
In Section \ref{subsec:rq}, we introduce four research questions, followed by detailing the case studies in Section \ref{subsec:case}. Section \ref{subsec:metric} shows the evaluation metrics, while Section \ref{subsec:testing} introduces the statistical tests employed. Finally, we introduce the settings and execution environment in Section \ref{subsec:parameter}.

\subsection{Research Questions (RQs)}
\label{subsec:rq}

 \vspace{-1mm}
In this paper, we plan to answer four RQs as follows.
\begin{itemize}
    \item \textbf{RQ1}: How effective is \method in \sa, as compared to the baselines?
    \item \textbf{RQ2}: How efficient and effective is transfer learning?
    \item \textbf{RQ3}: Does UQ help to improve the performance of transfer learning? If so, which UQ method is the best?
    \item \textbf{RQ4}: Is prompt tuning effective and efficient for improving the performance of transfer learning?
\end{itemize}
\textcolor{black}{RQ1 comparatively analyzes \method against two baseline solutions: RISE-DT and a neural network model referred to as NN. RISE-DT has previously achieved state-of-the-art results in our prior research. Considering both \method and RISE-DT are DT-based solutions, we further compare \method with a non-DT machine learning baseline to demonstrate the advantages of \method. To the best of our knowledge, however, there are no other existing machine learning approaches applied to the elevator and ADS case studies. Therefore, we design the baseline NN by removing DTM from PPT, resulting in a combination of transformer layers, GRU layers, and CNN layers, which are well-established and widely utilized in the literature.}. RQ2-RQ4 dissect \method and assess the cost-effectiveness of introducing transfer learning, UQ, and prompt tuning to it. Specifically, RQ2 evaluates the efficiency and effectiveness of transfer learning by comparing the performance of \method with/without transfer learning (denoted as w/o TL). \textcolor{black}{\method w/o TL is trained with only data from the target domain while \method employs transfer learning to take advantage of data from the source domain as well.}  With RQ3, we study the impact of using or not using UQ on the performance of \method (denoted as w/o UQ) and select the most suitable UQ method from three UQ methods: CS score, Bayesian method and ensemble methods (denoted as CS, BUQ and EUQ, respectively). With RQ4, we plan to assess the improvement brought about by prompt tuning by comparing \method with and without prompt tuning (denoted as w/o PT).

\subsection{Case Studies}
\label{subsec:case}
\subsubsection{Orona Elevator System}
\label{subsubsec:orona}
\textbf{Orona elevator system} was studied in our prior work, where we evaluate RISE-DT with 11 versions dispatchers $d_*,d_1,d_2,...,d_{10}$, and two traffic templates, i.e., Lunchpeak $\Gamma_L$ and Uppeak traffic templates $\Gamma_U$. Notice the dispatcher $d*$ denotes the best dispatcher and $d_{1:10}$ denotes ten previous versions. In this work, we acquire 20 more dispatchers $d_{11},d_{12},...,d_{30}$ for a more comprehensive evaluation of \method. In total, we have access to 62 different subject systems $\Sigma^E_1=\langle d_{*},  \Gamma_L\rangle,\Sigma^E_2=\langle d_{*},  \Gamma_L\rangle,\Sigma^E_3=\langle d_1,  \Gamma_L\rangle,...,\Sigma^E_{32}=\langle d_{30}, \Gamma_L\rangle,\Sigma^E_{33}=\langle d_1,  \Gamma_U\rangle,...,\Sigma^E_{62}=\langle d_{30}, \Gamma_U\rangle$.
We collect 62 datasets $\mathcal{D}_1,.., \mathcal{D}_{62}$ from these subject systems by performing simulation on Elevate, a commercial simulator used by Orona to test their dispatchers in software in the loop simulation environment. These 62 subject systems can be categorized into 
four types of subject systems: \textit{LunchBest} (or \textit{LunchWorse}) denoting that the elevator dispatcher with the highest performance (or sub-par) operates during lunch rush (12:15 - 13:15 p.m.); \textit{UpBest} (or \textit{UpWorse}) representing that the best (or an under-performing) elevator dispatcher operates during the morning rush hour (8:30 - 9:30 a.m.). \textcolor{black}{ The \sa in this case study takes passenger attributes as input and predicts the waiting time before the elevator arrives. Passenger attributes include the time each passenger arrives at a floor, the passenger's mass, the time a passenger takes to get inside/outside the elevator, and arrival and destination floors. Further details on these attributes can be consulted in~\cite{han_uncertainty-aware_2023}.  }

\noindent\textbf{Evolution dataset construction}, in this case study, encompasses four types of evolutions: (1) $LunchBest\rightarrow UpBest$;(2) $UpBest\rightarrow LunchBest$; (3) $LunchWorse\rightarrow LunchBest$;  (4) $UpWorse\rightarrow UpBest$. (1) and (2) are dispatcher-variant evolutions where the dispatcher has undergone changes in the evolution. (3) and (4) are traffic-variant evolutions, where the source and target subject systems only differ in traffic templates, while the elevator dispatcher remains unchanged.

 \subsubsection{Autonomous Driving Systems Dataset}
\textbf{ADS dataset} is taken from DeepScenario \cite{c_lu_deepscenario_2023}-- an open-source dataset containing 33530 driving scenarios, generated by executing the reinforcement learning method DeepCollision with various test settings ~\cite{lu_learning_2023}. These scenarios were generated with different strategies (e.g., greedy search and reinforcement learning) to achieve various objectives (e.g., reducing time-to-collision and distance to obstacles). Each driving scenario is characterized by the properties and behaviors of the self-driving car under study (known as the ego vehicle) and other objects in the driving environment, such as pedestrians and other cars (known as NPC vehicles). One example can be described as "A red BoxTruck is overtaking the ego vehicle and maintaining the lane." DeepScenario provides us with 19 features of the ego and NPC vehicles about their speed, location, rotation, etc.
The complexity of driving scenarios differs regarding the number of NPCs involved. For example, a driving scenario without any NPC vehicle is much less challenging for the ADS of the ego vehicle to make decisions as compared to a scenario with NPC vehicles around. We acquired two datasets from DeepScenario with different complexity levels. We name the dataset with fewer NPC vehicles on average as the \textit{Simple} dataset and the one with more NPC vehicles as the \textit{Complex} dataset. Their descriptive statistics are given in Table \ref{tab:ads_data}. \textcolor{black}{Samples in these datasets comprise the vehicle's properties, including dynamic properties (e.g., rotation, velocity, and angular velocity) and geographic properties (e.g., GPS and position)~\cite{c_lu_deepscenario_2023}. The \sa takes these properties as input and aims to predict the time to collisions.}

\begin{table}[htb]
    \centering
    \scriptsize
    \caption{Descriptive statistics of the number of NPC vehicles in the \textit{Simple} and \textit{Difficult} datasets. Q1, Q2 and Q3 denote 25\%, 50\% and 75\% quantiles.}

    \begin{tabular}{c|ccccccc}
    
    \toprule
         \thead{Dataset} & \thead{Mean} & \thead{Std} & \thead{Min} &\thead{Q1}& \thead{Q2}& \thead{Q3} & \thead{Max}  \\
         \midrule
         Easy & 4.81 &3.59 & 0 & 2 & 4 & 7 & 21 \\
         Difficult &  6.51 & 3.96 &  0 &  3 &  6 &  9 &  36 \\
         \bottomrule
    \end{tabular}
    
    \label{tab:ads_data}
\end{table}

\noindent\textbf{ADS evolution dataset construction} encompasses the bidirectional evolution between the \textit{Simple} and \textit{Complex} datasets, i.e., $Simple\rightarrow Complex$ and $Complex \rightarrow Simple$.
 
\subsection{Metrics}
\label{subsec:metric}
We introduce the metrics to evaluate the predictive performance in Section \ref{subsubsec:ppem} and efficiency metrics in Section \ref{subsubsec:epem}. Metrics for assessing the UQ methods are introduced in Section \ref{subsubsec:uqem}. In Section \ref{subsec:testing}, we present the statistical tests used in the evaluation.

\subsubsection{Predicative Performance Evaluation Metrics}
\label{subsubsec:ppem}
\sa is essentially a regression prediction task. In this study, we prefer to use Huber loss because, unlike Mean Squared Error (MSE), Huber loss does not heavily penalize data points that deviate significantly from the rest, thus making the prediction model more robust in handling outliers in the data. Huber loss is calculated using the following formula~\cite{meyer_alternative_2021}, which involves two conditions:
\begin{equation}
L_{\delta}(y,f(x))=
\begin{cases}
\frac{1}{2}(y - f(x))^2 & \text{for } |y - f(x)| \leq \delta, \\
\delta \cdot |y - f(x)| - \frac{1}{2}\delta^2 & \text{otherwise.}
\end{cases}
\end{equation}
\noindent In this equation, $y$ is the true value, $f(x)$ is the predicted value, and $\delta$ is a hyperparameter that controls the transition between the loss for small and large residuals.

\subsubsection{Efficiency Performance Evaluation Metrics}
\label{subsubsec:epem}
We evaluate the efficiency with training time spent by \method's pretraining and prompt tuning phases. We denote $\mathcal{D}^S$ and $\mathcal{D}^T$ as the source and target datasets for transfer learning. \method's pretraining is executed on $\mathcal{D}^{pre}=\{\langle \mathcal{D}^{SPre}_1,\mathcal{D}^{TPre}_1  \rangle,\langle \mathcal{D}^{SPre}_2,\mathcal{D}^{TPre}_2  \rangle,...,\langle \mathcal{D}^{SPre}_N,\mathcal{D}^{TPre}_N \rangle\}$ of $N$ individual transfers. We determine the convergence time for one transfer with Equation \ref{eq:convergence}, where $time_{early\_stopping\_end}$ denotes the point at which early stopping occurs (i.e., no improvement for five consecutive epochs), while $time_{start}$ signifies the commencement point of training. 
\begin{equation}
    \label{eq:convergence}
    time_{convergence}=time_{early\_stopping\_end}-time_{start}
\end{equation}

\noindent\textit{Pretraining time} is computed by aggregating the convergence times for $N$ individual transfers with Equation \ref{eq:pretrain}.
\begin{equation}
    \label{eq:pretrain}
    time_{pretrain}=\sum_{i=1}^{N} time_{convergence} (\mathcal{D}^{SPre}_i, \mathcal{D}^{TPre}_i)
\end{equation}
\textit{Prompt tuning time} is determined by the convergence time on source dataset $\mathcal{D}_S$ and target dataset $\mathcal{D}_T$, as defined by Equation \ref{eq:finetune}.
\begin{equation}
    \label{eq:finetune}
    time_{prompttuning}=time_{convergence} (\mathcal{D}^S, \mathcal{D}^T)
\end{equation}

\subsubsection{UQ Method Evaluation Metrics}
\label{subsubsec:uqem}
\textbf{UQ Effectiveness Metric.} We compare samples selected by each UQ method with Precision@K; Let $l_A$ and $l_B$ denote the samples selected by method A and method B, respectively and \textit{Precision@K} measures to what extent $l_A$ and $l_B$ overlap in the top $K$ samples (Equation \ref{eq:pk}).

\begin{equation}
    \label{eq:pk}
    \text{Precision@K} = \frac{overlap(l_A, l_B)}{K}
\end{equation}

\noindent\textbf{UQ Efficiency Metric} measures the efficiency of a UQ method as the total time $\tau_{UQ}$ required for sample selection. We denote $\tau_{UQ}$ required by CS score, Bayesian and ensemble UQ as $\tau_{CS}$, $\tau_{BUQ}$ and $\tau_{EUQ}$, respectively. 

\subsection{Statistical testing}\label{subsec:testing}

To counteract the inherent variability associated with training neural networks, we conducted each experiment 30 times. Subsequently, we employed the Mann-Whitney U test~\cite{arcuri_practical_2011} to investigate the statistical significance of observed improvements of \method over the baseline RISE-DT. This was done for all pair-wise comparisons within each RQ. The baseline assumption or null hypothesis presumes no significant distinction between \method and RISE-DT under comparison. If this null hypothesis is dispelled, we deduce that they are not equivalent. We choose the significance level as 0.01; thus, $p-value<0.01$ denotes a significant improvement $\Delta$.

As recommended in~\cite{arcuri_practical_2011}, we chose Vargha and Delaney's A12 as the measure of effect size. This metric illustrates the probability of \textit{\method} outperforming \textit{RISE-DT}. If the A12 value exceeds 0.5, we can infer that \textit{\method} is more likely to yield superior results compared to \textit{RISE-DT}, and vice versa. We consider the effect size in the range  $[0.56, 0.64)$ as \emph{Small} $\downarrow$,  $[0.64, 0.71)$ as \emph{Medium} $\rightarrow$, and  $[0.71, 1]$ as \emph{Large} $\uparrow$.

\subsection{Settings and Execution}
\label{subsec:parameter}
Assigning hyperparameter values manually can potentially introduce bias. To mitigate this issue, we carried out a 10-fold cross-validation process to select optimal hyperparameters. This involved partitioning the dataset into 10 sequential segments, using the first nine for training and the final one for validation. Due to the difference in complexity, we set the hyperparameters differently for the elevator and ADS subject systems. We present some key values in Table \ref{tab:hyper}
\begin{table}[hbt]
\scriptsize
    \centering
    \caption{Hyperparameter values for \method. d\_model, n\_heads, dim\_feedforward, n\_layers denote the hidden dimension, number of heads, feedforward network dimension, and number of the MHSA modules in the transformer. proj\_dim represents the dimension of the projection module in the transfer learning component.}
    \begin{tabular}{c|c|c}
    \toprule
        \thead{Parameter}  & \thead{For Elevator System} & \thead{For ADS}\\
    \midrule
    \cellcolor{lightgray}d\_model &\cellcolor{lightgray}16 &\cellcolor{lightgray}128 \\
    batch size &1 &1 \\
    \cellcolor{lightgray}n\_heads &\cellcolor{lightgray}1 &\cellcolor{lightgray}32 \\
    dim\_feedforward &128 &1024 \\
    \cellcolor{lightgray}n\_layers&\cellcolor{lightgray}1 & \cellcolor{lightgray}24\\
    proj\_dim & 32& 128\\
    \bottomrule
    \end{tabular}
    
    \label{tab:hyper}
\end{table}
 
Our code is written in Python with Pytorch 2.0 library~\cite{paszke_pytorch_2019}. \cs is calculated with Uncertainty Toolbox~\cite{chung_uncertainty_2021}. We execute our code on a national, experimental, heterogeneous computational cluster called eX3. This node contains 2x Intel Xeon Platinum 8186, 1x NVIDIA V100 GPUs. 

\section{Results and Analysis}
In this section, we answer each RQ. A replication package of \method is provided here for reference \footnote{https://github.com/qhml/ppt}.
\label{sec:experiment_results}
\subsection{RQ1 - \method's Overall Effectiveness}
\label{subsec:rq1_results}

\textcolor{black}{RQ1 aims to evaluate the overall effectiveness of \method in TTE analysis by comparing it to two baselines: RISE-DT and NN. Table \ref{tab:rq1} presents the results of both Orona's elevator and DeepScenario's ADS case studies.} \textcolor{black}{Compared to \textbf{RISE-DT} in the \textit{elevator case study}, we find that \method demonstrates superiority in both traffic-variant evolutions (i.e., $UpBest\rightarrow LunchBest$ and $LunchBest\rightarrow UpBest$) and dispatcher-variant evolutions i.e., $LunchWorse\rightarrow LunchBest$ and $UpWorse\rightarrow UpBest$). The minimum improvement is 5.70 (109.49-103.79), for the case of $LunchBest\rightarrow UpBest$), while the maximum improvement is 9.62 for the case of $UpWorse\rightarrow UpBest$. } According to the statistical testing results, we find that the improvements in the traffic-variant evolutions are significant ($p-value < 0.01$) with strong effect sizes ($A12>0.71$). The majority of the improvements in the dispatcher-variant evolutions are significant (21 out of 30) and the effect sizes are mostly strong (17 out of 30 for $LunchWorse\rightarrow LunchBest$ and 20 out of 30 for the case $UpWorse\rightarrow UpBest$). \textcolor{black}{In the \textit{ADS case study}, \method outperforms RIST-DT by 14.410 and 10.75 for the Simple$\rightarrow$Complex and   Complex$\rightarrow$Simple cases respectively. we observe significance and very strong effect sizes (close to 1) in both improvements. } 

\textcolor{black}{Compared to \textbf{NN} in the \textit{elevator case study,} we find that PPT reaches much lower Huber loss in all cases, with a minimum reduction of 8.25 (91.2-82.95) for the $LunchWorse\rightarrow LunchBest$ case. Statistical testing results demonstrate the significance of all results and most effect sizes are strong for both traffic-variant (2 out of 2) and dispatcher-variant evolutions (20 out of 30 for $LunchWorse\rightarrow LunchBest$ and 15 out of 30 for $UpWorse\rightarrow UpBest$). In the \textit{ADS case study}, \method outperforms NN by 16.60 (232.45-215.85) and 15.99 (124.05-108.06) for Simple$\rightarrow$Complex and Complex$\rightarrow$Simple, respectively. Statistical testing shows the significance of all results and strong effect size (A12$=1$ for both cases). }
\begin{table*}[]
\center
\scriptsize
\caption{\textcolor{black}{Huber loss, Mann-Whitney statistical test results, and A12 effect sizes of comparing the baselines and \method (RQ1). $\Delta$ denotes a significant testing result; S, M, and L represent small, medium, and large effect sizes, respectively.}}
\label{tab:rq1}
\begin{tabular}{c|c|cccc}
\toprule
                           \thead{Dataset}& \thead{Evolution}                                         & \thead{Metric \& Testing}  & \thead{NN}                                                           & \thead{RISE-DT}                                                      & \thead{PPT}    \\ \midrule
\multirow{12}{*}{Orona Elevator} & \multirow{3}{*}{UpBest$\rightarrow$LunchBest}     & \cellcolor{darkgray}Huber  Loss & \cellcolor{darkgray}89.41                                                        & \cellcolor{darkgray}87.89                                                         & \cellcolor{darkgray}80.14  \\
                           &                                                   & p-value & $\Delta$                                                     & $\Delta$                                                      & -      \\
                           &                                                   & A12     & 1.00(L)                                             & 0.828 (L)                                            & -      \\\cline{2-6}
                           & \multirow{3}{*}{LunchBest$\rightarrow$UpBest}     & \cellcolor{darkgray}Huber Loss  & \cellcolor{darkgray}115.74                                                       & \cellcolor{darkgray}109.49                                                        & \cellcolor{darkgray}103.79 \\
                           &                                                   & p-value & $\Delta$                                                     & $\Delta$                                                      & -      \\
                           &                                                   & A12     & 1.00(L)                                             & 0.774 (L)                                            & -      \\\cline{2-6}
                           & \multirow{3}{*}{LunchWorse$\rightarrow$LunchBest} & \cellcolor{darkgray}Huber Loss  & \cellcolor{darkgray}91.20                                                        & \cellcolor{darkgray}89.11                                                         & \cellcolor{darkgray}82.95  \\
                           &                                                   & p-value & $\Delta\times30$                                             & $\Delta\times 21$                                             & -      \\
                           &                                                   & A12     & S$\times 2; $M$\times 8; $L$ \times 20$ & S$\times 4; $M$\times 3; $L$ \times 17$ & -      \\\cline{2-6}
                           & \multirow{3}{*}{UpWorse$\rightarrow$UpBest}       & \cellcolor{darkgray}Huber Loss  & \cellcolor{darkgray}117.36                                                       &\cellcolor{darkgray} 114.88                                                        & \cellcolor{darkgray}105.26 \\
                           &                                                   & p-value & $\Delta\times 30$                                            & $\Delta\times 21$                                             & -      \\
                           &                                                   & A12     & S$\times5;$ M$\times 10; $L $\times 15$ & S$\times 2;$ M$ \times 1; $L$ \times 20$ & -      \\\midrule
\multirow{6}{*}{DeepScenario ADS}       & \multirow{3}{*}{Simple$\rightarrow$Complex}       & \cellcolor{darkgray}Huber Loss  & \cellcolor{darkgray}232.45                                                       & \cellcolor{darkgray}230.26                                                        & \cellcolor{darkgray}215.85 \\
                           &                                                   & p-value & $\Delta$                                                     & $\Delta$                                                      & -      \\
                           &                                                   & A12     & 1.00(L)                                             & 0.993 (L)                                            & -      \\\cline{2-6}
                           & \multirow{3}{*}{Complex$\rightarrow$Simple}       & \cellcolor{darkgray}Huber Loss  & \cellcolor{darkgray}124.05                                                       & \cellcolor{darkgray}118.81                                                        & \cellcolor{darkgray}108.06 \\
                           &                                                   & p-value & $\Delta$                                                     & $\Delta$                                                      & -      \\
                           &                                                   & A12     & 1.00(L)                                             & 0.927 (L)                                            & -      \\ \bottomrule
\end{tabular}
\end{table*}

 \noindent\cbox[gray!30]{
We conclude that \method is effective in \sa in both elevator and ADS case studies, \textcolor{black}{significantly outperforming both baselines: RISE-DT and NN, on average with large effect sizes in both case studies.}}

 \subsection{RQ2 - Transfer Learning Effectiveness and Efficiency}
 With RQ2, we aim to investigate the contribution of the transfer learning in \method. Table \ref{tab:results_ablation}  presents the experiment results of comparing \method and \method without transfer learning (denoted as "w/o TL"). We find that the Huber loss increases sharply after removing transfer learning. Specifically, in the elevator case study, the average increase reaches 21.32, with a minimum increase of 12.96 for the case $LunchWorse\rightarrow LunchBest$. In the ADS case study, the increases in Huber loss are even higher, with an average value of 28.26.
  \begin{table*}[hbt]
     \caption{Huber loss results of the ablation study (RQ2-RQ4). 
     "w/o TL", "w/o UQ" and "w/o PT" represent \method without transfer learning, without UQ, and without prompt tuning, respectively; "CS", "BUQ", and "EUQ" denote \cs, Bayesian UQ, and Ensemble UQ, respectively.}
\label{tab:results_ablation}
     \centering
     \scriptsize
     \begin{tabular}{c|c|c|cccc|c|c}
     \toprule
          & \thead{Evolution} &\thead{w/o TL }& \thead{w/o UQ}& \thead{CS} & \thead{BUQ} &\thead{EUQ}  & \thead{w/o PT} & \thead{PPT} \\
     \midrule
\multirow{5}{*}{Orona Elevator} &  
\cellcolor{lightgray} UpBest$\rightarrow$LunchBest&\cellcolor{lightgray}  95.91 &\cellcolor{lightgray} 87.56 &\cellcolor{lightgray} 80.14&\cellcolor{lightgray} 85.23&\cellcolor{lightgray} 79.40&\cellcolor{lightgray} 86.31 & \cellcolor{lightgray} 80.14  \\
         &LunchBest$\rightarrow$UpBest& 132.80 &105.09 &103.79 &101.80 &105.20 &109.07 & 103.79  \\
         &\cellcolor{lightgray} LunchWorse$\rightarrow$LunchBest&\cellcolor{lightgray}  95.91&\cellcolor{lightgray} 87.04 &\cellcolor{lightgray} 82.95&\cellcolor{lightgray} 85.79&\cellcolor{lightgray} 80.09 &\cellcolor{lightgray} 88.65& \cellcolor{lightgray} 82.95  \\
         &UpWorse$\rightarrow$UpBest& 132.80&108.76&105.26&113.28&103.67&107.68 & 105.26 \\
         &\cellcolor{darkgray}Average  & \cellcolor{darkgray}114.35  & \cellcolor{darkgray}97.11 & \cellcolor{darkgray}93.4&\cellcolor{darkgray}96.51&\cellcolor{darkgray}92.07&\cellcolor{darkgray}97.93 &\cellcolor{darkgray} 93.03\\ \hline
         \multirow{3}{*}{DeepScenario ADS} &\cellcolor{lightgray} Simple$\rightarrow$Complex&\cellcolor{lightgray}  241.19 &\cellcolor{lightgray} 224.80&\cellcolor{lightgray} 215.85&\cellcolor{lightgray} 219.44&\cellcolor{lightgray} 216.10&\cellcolor{lightgray} 219.25 &\cellcolor{lightgray}  215.85   \\
         &Complex$\rightarrow$Simple& 139.23&115.25&108.06&112.90&110.60&110.94 & 108.06  \\
         &\cellcolor{darkgray} Average & \cellcolor{darkgray}190.21&\cellcolor{darkgray}170.02 & \cellcolor{darkgray}161.95&  \cellcolor{darkgray}166.15&\cellcolor{darkgray}163.35 & \cellcolor{darkgray}165.10 & \cellcolor{darkgray}161.95 \\
    \bottomrule
     \end{tabular}
 \end{table*}

We also investigated the efficiency of transfer learning as depicted in Table \ref{tab:tl_time}. In this table, we report the training time of transfer learning, comprising the pretraining phase (denoted as "Pre") and prompt tuning phase (denoted as "PT"). We find that pretraining consumes marginally more time compared to prompt tuning. The average pretraining times in the elevator and ADS case study are 70 hours and 97.5 hours, respectively. Whereas the prompt tuning time in these two cases is merely 2.77 hours and 5.05 hours, respectively. Such results are expected since the pretraining dataset is larger than the prompt tuning dataset. Moreover, the pretraining phase only requires a single execution before the transfer learning, making the large pretraining time acceptable for the production environment.

\begin{table}[hbt]
    \centering
    \caption{Time cost of each training phase in \method. "Pre", "PT" and "Total" denote the time cost for the pretraining phase, prompt tuning phase, and sum of both.}
    \resizebox{\columnwidth}{!}{
    \begin{tabular}{c|cccc}
    \toprule
        &\multicolumn{1}{c}{\thead{Evolution}}&\thead{Pre} & \thead{PT} & \thead{Total}\\
        \midrule
        \multirow{4}{*}{Elevator}& \cellcolor{lightgray}UpBest$\rightarrow$LunchBest&\cellcolor{lightgray} 76h &\cellcolor{lightgray} 2.9h &\cellcolor{lightgray} 78.9h \\
         &LunchBest$\rightarrow$UpBest&76h & 4.1h & 80.1h\\
         &\cellcolor{lightgray}LunchWorse$\rightarrow$LunchBest&\cellcolor{lightgray}61h & \cellcolor{lightgray}2.5h & \cellcolor{lightgray}63.5h\\
         &UpWorse$\rightarrow$UpBest&67h & 1.6h & 68.6h\\
         &\cellcolor{darkgray}Average  & \cellcolor{darkgray}70h & \cellcolor{darkgray}2.77h & \cellcolor{darkgray}72.77h \\\hline
         \multirow{2}{*}{ADS} &\cellcolor{lightgray}Simple$\rightarrow$Complex& \cellcolor{lightgray}98h & \cellcolor{lightgray}5.1h & \cellcolor{lightgray}103.1h\\
         &Complex$\rightarrow$Simple& 97h & 5.0h & 102h\\
         &\cellcolor{darkgray}Average & \cellcolor{darkgray}97.5h & \cellcolor{darkgray}5.05h & \cellcolor{darkgray}102.05h\\
        \bottomrule
    \end{tabular}
    }
    \label{tab:tl_time}
\end{table}

\noindent\cbox[gray!30]{
We conclude that transfer learning is effective in both elevator and ADS case studies. Removing transfer learning from \method leads to surges in Huber loss, i.e., 21.32 and 28.26 hours in the elevator and ADS case studies. The majority of the time cost of transfer learning is spent in the pretraining phase, which we believe acceptable as one only needs to pretrain \method once. 
}

\subsection{RQ3 - UQ Effectiveness and efficiency}
RQ3 aims to evaluate the effectiveness and efficiency of UQ by comprehensively assessing its influence on \sa, selected samples, and time cost. 

\noindent\textbf{UQ's Influence on TTE Analysis.} To highlight the effectiveness of UQ, we compare \method and \method without UQ (denoted as "w/o UQ") in Table \ref{tab:results_ablation}. In the elevator case study, we see an average increase of 4.08 after removing UQ from \method. The maximum increase is 7.42 for case $UpBest\rightarrow LunchBest$. In the ADS case study, we find the Huber loss boost from 215.85 to 224.80 for case $Simple \rightarrow Complex$ and from 108.06 to 115.25 for case $Complex \rightarrow Simple$. 

We also compare three UQ methods (i.e., \cs, Bayesian and Ensemble UQ, denoted as CS, BUQ and EUQ) in terms of Huber loss in Table \ref{tab:results_ablation}. In the elevator case study, we find EUQ tends to be the most effective UQ method, achieving the lowest Huber loss in all evolutions except for $LunchBest\rightarrow UpBest$. However, the difference between CS and EUQ is nominal with a maximum of 1.59 (105.26-103.67) for $UpWorse \rightarrow UpBest$. In the ADS case study, CS beats EUQ and shows the lowest Huber loss for both evolutions: $Simple \rightarrow Complex$ and $Complex \rightarrow Simple$.
    

\noindent\textbf{UQ's Influence on Samples Selected.} To compare the three UQ methods, we look into samples selected by the UQ methods. We calculate the Precision@K metrics to demonstrate the overlaps among the methods and the results are shown in Table \ref{tab:uq_pk}. The precision@1 (denoted as P@1) and precision@3 (denoted as P@3) are all 100\% in each evolution in the elevator and ADS case studies, indicating that the top 1 and top 3 samples selected by one UQ method are always selected by the other two methods. The precision@10 metric (denoted as P@10) gives lower results. In the elevator case study, the precision@10 metric remains 100\% in the traffic-variant evolutions (i.e., $UpBest\rightarrow LunchBest$ and $LunchBest \rightarrow UpBest$), indicating the top 10 samples selected by one UQ method are also selected by the other two. As for the dispatcher-variant evolutions, the precision@10 results are still high ($\geq0.93$), though not 100\%. In the ADS case study, the precision@10 results are higher than 82\%, implying approximately 8 out of the top 10 samples selected by one UQ method are also selected by the other two.

\begin{table}[hbt]
    \centering
    \caption{Results of Precision@K (abbreviated as P@K) for top K samples selected by the UQ methods, where $K={1, 3, 10}$.}
    \resizebox{\columnwidth}{!}{
    \begin{tabular}{c|cccc}
    \toprule
        &\multicolumn{1}{c}{\thead{Evolution}} & \thead{P@1} & \thead{P@3} & \thead{P@10} \\
        \midrule
        \multirow{4}{*}{Elevator} &  \cellcolor{lightgray} UpBest$\rightarrow$ \cellcolor{lightgray}LunchBest& \cellcolor{lightgray}100\% & \cellcolor{lightgray}100\% &\cellcolor{lightgray}100\% \\
         &LunchBest$\rightarrow$UpBest& 100\% & 100\%&100\%\\
         &\cellcolor{lightgray}LunchWorse$\rightarrow$LunchBest&\cellcolor{lightgray}100\%&\cellcolor{lightgray}100\%&\cellcolor{lightgray}95\%\\
         &UpWorse$\rightarrow$UpBest&100\%&100\%&93\% \\\hline
         \multirow{2}{*}{ADS} &\cellcolor{lightgray}Simple$\rightarrow$Complex&\cellcolor{lightgray}100\% &\cellcolor{lightgray}100\% &\cellcolor{lightgray}82\%\\
         &Complex$\rightarrow$Simple&100\% & 100\% &86\%\\
         \bottomrule
    \end{tabular}
    }
    \label{tab:uq_pk}
\end{table}

\noindent\textbf{UQ's Time Cost.} To assess the efficiency of UQ methods, we calculate the time cost for each UQ method in the elevator and ADS case studies as in Table \ref{tab:uq_time}. We find that the time cost for ADS (127.04 seconds on average) is much higher than that for the elevator case study (56.40 seconds on average). \cs spends the least time (denoted as $\tau_{CS}$), while the ensemble UQ method (denoted as $\tau_{EUQ}$) takes the most time to perform UQ in both case studies.
 
\begin{table}[hbt]
    \centering
    \caption{Time cost of the three  UQ methods. $\tau_{CS}$, $\tau_{BUQ}$ and $\tau_{EUQ}$ denote the time cost for CS score, Bayesian and ensemble UQ methods. }
   \resizebox{\columnwidth}{!}{
   \begin{tabular}{c|ccc|c}
\toprule
   Case Study & $\tau_{CS}$ & $\tau_{BUQ}$ & $\tau_{EUQ}$ & Average \\
    \midrule
    Elevator System& \textbf{10.27s} & 77.00s & 81.92s & 56.40s \\
     ADS&\textbf{25.81s} &156.02s &199.3s & 127.04s \\
     \bottomrule
\end{tabular}
    }
    \label{tab:uq_time}
\end{table}
\noindent\cbox[gray!30]{
\textcolor{black}{We conclude that ensemble UQ demonstrates the highest effectiveness while consuming much longer time. \cs is most efficient compared to ensemble and Bayesian UQ while retaining comparable effectiveness as ensemble UQ in \sa. Different UQ methods should be chosen for various case studies, e.g., CS score should be preferred when resources are limited, while ensemble UQ is recommended for cases of prioritizing effectiveness.}
}
\label{subsec:rq2_results}

\subsection{RQ4 - Prompt Tuning Effectiveness and Efficiency}
\label{subsec:rq3_results}
RQ4 aims to demonstrate the effectiveness and efficiency of prompt tuning. We first compare \method with \method without prompt tuning (denoted as "w/o" PT) as shown in Table \ref{tab:results_ablation}. In the elevator case study, \method outperforms \method without prompt tuning by 4.89 on average. The maximum Huber loss reduction is 6.17 for the case $UpBest \rightarrow LunchBest$. In the ADS case study, we find similar reductions in both evolutions of $Simple \rightarrow Complex$ and $Complex \rightarrow Simple$ with an average of 3.14.


As for efficiency, we compare prompt tuning (denoted as PT) with fine-tuning (denoted as FT) and report the time cost in Table \ref{tab:pt_time}. Notice that fine-tuning is used in RISE-DT. We find both fine-tuning and prompt tuning times in the ADS case study are higher than in the elevator case study. However, there is no dominating choice in terms of time cost since the time spent on fine-tuning and prompt tuning is quite close. Fine-tuning takes less time for cases $UpBest\rightarrow LunchBest$ and 
$LunchWorse\rightarrow LunchBest$ in the elevator case study and $Complex\rightarrow Simple$ in the ADS case study. 

\begin{table}[hbt]
    \centering
    \caption{Time cost of fine tuning and prompt tuning. Column "Difference" shows the difference between the time cost of fine-tuning and prompt tuning.}
    \resizebox{\columnwidth}{!}{
    \begin{tabular}{c|cccc}
    \toprule
        &\multicolumn{1}{c}{\thead{Evolution}}& \thead{FT} & \thead{PT} & \thead{Difference} \\
        \midrule
        \multirow{4}{*}{Elevator}&\cellcolor{lightgray} UpBest$\rightarrow$LunchBest& \cellcolor{lightgray} 2.7h &\cellcolor{lightgray}  2.9h & \cellcolor{lightgray} -0.2h \\
         &LunchBest$\rightarrow$UpBest&4.1h & 4.1h & 0h \\
         &\cellcolor{lightgray} LunchWorse$\rightarrow$LunchBest& \cellcolor{lightgray} 2.4h & \cellcolor{lightgray} 2.5h & \cellcolor{lightgray} -0.1h \\
         &UpWorse$\rightarrow$UpBest&1.8h & 1.6h & 0.2h \\
         &\cellcolor{darkgray} Average & \cellcolor{darkgray}2.75h & \cellcolor{darkgray}2.77h & \cellcolor{darkgray}-0.02h \\\hline
         \multirow{2}{*}{ADS} &\cellcolor{lightgray}Simple$\rightarrow$Complex& \cellcolor{lightgray}5.3h & \cellcolor{lightgray}5.1h & \cellcolor{lightgray}0.2h \\
         &Complex$\rightarrow$Simple& 4.9h & 5.0h & -0.1h \\
         &\cellcolor{darkgray}Average  & \cellcolor{darkgray}5.1h & \cellcolor{darkgray}5.05h & \cellcolor{darkgray}0.1h \\
        \bottomrule
    \end{tabular}
    }
    \label{tab:pt_time}
\end{table}
\noindent\cbox[gray!30]{We conclude that prompt tuning effectively reduces Huber loss in \sa, and the time cost is approximately on the same level as fine-tuning. }

\subsection{Threats to Validity}

\noindent\textbf{Construct Validity} concerns if the metrics we used can reflect the quality of TTE analysis. We are aware of other metrics such as MSE and RMSE. However, choosing the Huber loss, instead of MSE or RMSE, is because the Huber loss is more robust against outliers than MSE and RMSE, hence providing a more stable evaluation of \method.
\noindent\textbf{Internal Validity} refers to the extent to which the cause-and-effect relationship aiming to be established in our study is not due to other factors. \textcolor{black}{One possible threat lies in the choice of hyperparameters (e.g., number of neurons, learning rate). To alleviate this issue, we performed a 10-fold cross-validation to select the hyperparameters automatically. }
\noindent Regarding \noindent\textbf{conclusion validity}, since \method is neural network-based and tends to introduce randomness into the experiments, to reduce the influence of randomness, we repeated each experiment 30 times, as suggested in \cite{arcuri_practical_2011}) and performed statistical testing to conclude the significance.
\noindent\textbf{External Validity} is about to what extent \method can generalize to other domains.\textcolor{black}{We design \method to be applicable for TTE analysis in CPSs. We are aware that \method can perform differently on different datasets. That is why we picked two datasets from different domains with different levels of complexity to provide more comprehensive evaluation results. } 

\section{Related work}\label{sec:related}
We discuss CPS safety and security (Section \ref{subsec:cps}), DT in CPSs (Section \ref{subsec:dt_in_cps}), transfer learning (Section~\ref{subsec:transfer}), UQ (Section~\ref{subsec:uq}) and prompt tuning (Section ~\ref{subsec:prompt_learning}).

\subsection{Cyber-physical Systems Security and Safety}\label{subsec:cps}
CPSs have risks originating from both physical and cyber dimensions. Thus, numerous security and safety methodologies have been proposed \cite{humayed_cyber-physical_2017, lv_levenbergmarquardt_2017}. 
With the growing deep learning (DL) application for enhancing CPS security and safety \cite{wickramasinghe_generalization_2018, lee_building_2019}, a critical bottleneck experienced by researchers and practitioners is the high cost of collecting data and, in some cases, the infeasibility of obtaining labeled data for real-world CPSs. This shortage of labeled data forms a roadblock for the effective training of DL models, which often require sufficient data from the system - a condition that cannot always be ensured. To combat this, we propose \method, a DL methodology tailored to tackle the scarcity of labeled CPS data by incorporating transfer learning and prompt tuning.

\subsection{Digital Twins in Cyber-physical Systems}
\label{subsec:dt_in_cps}
DT technologies facilitate real-time synchronization with CPSs \cite{eckhart_securing_2018}. For instance, DTs \cite{becue_cyberfactory1_2018} were used to analyze the appropriate engineering of CPSs under attack scenarios. Rules were incorporated into DTs in \cite{eckhart_towards_2018} to determine whether an attacker could compromise programmable logic controllers. It was recommended in \cite{bitton_deriving_2018} to test DT as a safer alternative to testing real CPS. Also, DTs were employed in \cite{damjanovic-behrendt_digital_2018} for the privacy assessment of actual smart car systems. Researchers also initiated the development of academic testbeds for DT development, such as the one described in ~\cite{wang_simplexity_2023}.
These examples show the advantages of DTs. However, our work is unique in its emphasis on the evolution and development of DTs.

\subsection{Transfer Learning}\label{subsec:transfer}
Transfer learning has four main strategies. The first is the \textit{model control strategy} implementing transfer learning at the model tier. For instance, the Domain Adaptation Machine (DAM) \cite{duan_learning_2012}  uses data from several source domains, constructs a classifier for each, and employs regularizers to maintain the final model's complexity. 
The second strategy, the \textit{parameter control strategy}, operates assuming that a model's parameters embody the knowledge it has assimilated. For instance, Zhuang et al. \cite{zhuang_open_2022} proposed directly sharing parameters between the source and target models for text classification.
The \textit{model ensemble strategy} is the third strategy, where transfer learning merges various source models. For instance, in \cite{gao_adaptive_2019}, several weak classifiers were trained with different model structures on multiple source domains and the final model is determined based on a weighted vote from these weak classifiers.
Lastly, \textit{deep learning transfer strategies} transfer knowledge between two deep learning models by aligning corresponding layers from source and target models. Zhuang et al. \cite{zhuang_open_2022} proposed a transfer learning method with an autoencoder that aligns reconstruction, distribution, and regression representations. 

In short, early strategies, e.g., model and parameter control, performed knowledge transfer with intuitive methods such as adding regularizers and sharing parameters. Their performance is comparable to the more recent model ensemble and deep learning transfer techniques. Model ensemble efficiently deals with multiple heterogeneous source domains ~\cite{farahani_brief_2020}, although it demands substantial computing resources. Deep learning transfer techniques transfer knowledge between two neural network models. Since \method is neural network-based, we follow this research trend, aligning the GRU and prediction layers representation.


\subsection{Uncertainty Quantification}\label{subsec:uq} 
Numerous UQ methods are derived from Bayesian methods. For example, probability theory was used to interpret neural network parameters in \cite{wang_adversarial_2018}. Later, Monte Carlo dropout was integrated into \cite{srivastava_dropout_nodate} as a regularization term for calculating prediction uncertainty, eliminating the need for posterior probability computation. Further, in \cite{salakhutdinov_bayesian_2008}, a stochastic gradient Markov chain Monte Carlo method was proposed requiring estimating the gradient on small mini-batch sets, significantly reducing computational load compared to direct posterior distribution estimation. 

Several open-source UQ tools exist. For instance, Uncertainty Wizard \cite{weiss_uncertainty_2022} is a TensorFlow Keras plugin supporting common UQ methods, including Bayesian and ensemble-based methods. Similarly, Uncertainty Toolbox \cite{chung_uncertainty_2021}, built on Pytorch, provides common Bayesian and ensemble UQ methods, alongside additional metrics such as calibration, sharpness, and accuracy.

UQ methods have been applied in various application domains. For instance, NIRVANA validates deep learning model predictions based on MC dropout \cite{catak_uncertainty-aware_2022}. Regarding uncertainty-aware analyses, Zhang et al. \cite{zhang_empirical_2018, zhang_machine_2020} proposed several methods for specifying, modeling, and quantifying uncertainties in CPSs and testing CPSs under uncertainties.

\subsection{Prompt Tuning}
\label{subsec:prompt_learning}
Prompt tuning has become popular recently. Multiple techniques have been proposed to enhance the effectiveness of prompt tuning. At the foundational level, several works have explored using prompts for language models. For example, GPT-3~\cite{brown_language_2020} uses prompts for natural language processing tasks without any explicit supervision. Inspired by this, AutoPrompt \cite{shin_autoprompt_2020} is proposed for automated prompt discovery for language models.
Several works are proposed for prompt selection, e.g., P-tuning \cite{liu_p-tuning_2022} incorporates trainable continuous prompts into pre-trained models showing impressive improvements on multiple benchmark datasets. Though prompt tuning is an exciting area with various strategies, toolkits, and applications being proposed, no prior work focused on using prompt tuning in DT construction and evolution. We developed our prompt-based learning method to fulfill the needs of our application domain.

\section{Conclusion and Future Work}\label{sec:conclusion}
We propose \method to evolve the DTs for Time-to-Event prediction in CPSs. To alleviate the data scarcity problem, we utilize transfer learning to transfer knowledge across different subject systems, with the help of uncertainty quantification and prompt tuning. We evaluate \method on two CPSs, namely an elevator system and an autonomous driving system (ADS). The experiment results show that \method is effective in \sa in both elevator and ADS case studies, averagely outperforming the baselines by 7.31 and 12.58, respectively, in terms of Huber loss. Further analysis into transfer learning, uncertainty quantification, and prompt tuning demonstrate their individual contribution to reducing the Huber loss. In the future, we plan to investigate more prompt tuning techniques by exploring other prompt designing methods. We are also interested in applying our method in other CPSs, such as power grids and railway systems, and evolving DTs alongside ADS internal functionality changes.


%

\section*{Acknowledgment}

Qinghua Xu is supported by the security project funded by the Norwegian Ministry of Education and Research. The work is also partially supported by the Horizon 2020 project ADEPTNESS (871319) funded by the European Commission and the Co-tester project (No. 314544) funded by the Research Council of Norway. The experiment has benefited from the Experimental Infrastructure for Exploration of Exascale Computing (eX3), which is financially supported by the Research Council of Norway under contract 270053.




%


\bibliographystyle{IEEEtran}
   \bibliography{references.bib}
%








\end{document}